# Strong spatial and spectral localization of surface plasmons in individual randomly disordered gold nanosponges

*Jinhui Zhong,[1,‡] Abbas Chimeh,[1,‡] Anke Korte,[1,‡] Felix Schwarz,[2] Juemin Yi,[1] Dong Wang,[3] Jinxin Zhan,[1] Peter Schaaf,[3] Erich Runge,[2] Christoph Lienau[1,\*]*

[1]Institut für Physik and Center of Interface Science, Carl von Ossietzky Universität, 26129 Oldenburg, Germany.

[2]Institut für Physik and Institut für Mikro- und Nanotechnologien MacroNano®, Technische Universität Ilmenau, 98693 Ilmenau, Germany.

[3]Institut für Mikro- und Nanotechnologien MacroNano® and Institut für Werkstofftechnik, Technische Universität Ilmenau, 98693 Ilmenau, Germany.

[4]Forschungszentrum Neurosensorik, Carl von Ossietzky Universität, 26111 Oldenburg, Germany.

ABSTRACT. Porous nanosponges, percolated with a three-dimensional network of 10-nm sized ligaments, recently emerged as promising substrates for plasmon-enhanced spectroscopy and (photo-)catalysis. Experimental and theoretical work suggests surface plasmon localization in some hot-spot modes as the physical origin of their unusual optical properties, but so far the existence of such hot-spots has not been proven. Here we use scattering-type scanning near-field nano-spectroscopy on individual gold nanosponges to reveal spatially and spectrally confined modes with 10 nanometer localization lengths by mapping the local optical density of states. High quality factors of individual hot-spots of more than 40 are demonstrated. A statistical analysis of near-field intensity fluctuations unveils plasmonics in the strong localization regime. The observed field localization and enhancement make such nanosponges an appealing platform for a variety of applications ranging from nonlinear optics to strong-coupling physics.

Nanoporous gold[1] has attracted considerable interest because of its unique three-dimensional bicontinuous ligament network and high surface-to-volume ratio, which benefit applications in sensing,[2] catalysis,[3-5] and supercapacitors.[6] Of particular interest are the plasmonic properties of this material since surface plasmon (SP) localization[7-9] in the randomly disordered ligament network favors, e.g., surface-enhanced Raman scattering (SERS)[10, 11] and infrared absorption[2] spectroscopy, and enhances fluorescence from embedded quantum emitters[12]. Recently, nanoporous gold in its nanoparticle form – "nanosponge" [13, 14] – has emerged as an promising plasmonic architecture.[15] Individual nanosponges with (sub-)micron diameters have been fabricated by firstly solid-state dewetting of a bimetallic film, such as gold/silver, to form an alloy nanoparticle. Subsequently, the alloy nanoparticle is dealloyed in which the less noble metal (silver) is chemically removed, leading to the formation of porous structure with nanometer-sized channels perforating the entire particle.[13, 14] Many of the structural parameters of these sponges like the particle and ligament sizes and porosity are highly tunable,[13] offering a means to tailor their plasmonic properties. Unlike conventional plasmonic antennas where Ohmic loss and radiative damping[16] dominate the decay of SP, the plasmon excitations of nanosponges may also experience multiple coherent scattering with the randomly disordered, nanoporous ligament network. Theoretical results predict that these multiple scattering processes can lead to the random localization of SP in certain hot-spots with small volumes.[17-19] The resulting substantial local field enhancement could make these particles attractive, for instance, for strong coupling to single quantum emitters,[20-25] applications in SERS[10, 11, 18] or enhancing non-linear optics.

SP localization has been studied quite extensively for quasi-two-dimensional (2D) percolated metal thin films.[7, 8] It is well known that multiple SP scattering leads to spectrally narrow-band SP modes with resonance linewidths down to a few tens of nm and mode areas of ~1000 $nm^2$, and enhancement of second harmonic generation has been demonstrated.[26] Much less is known, however, about SP localization in small three-dimensional (3D) nanosponges. Unlike 2D nanoporous films, nanosponges display a collective, dipolar SP resonance that can be excited by far-field light.[17, 19] This turns them into an efficient optical nanoantenna. Understanding the coupling between this collective mode and randomly localized SP may therefore reveal insight into the formation of plasmonic hot-spots and their coupling to radiative SP modes. While finite difference time domain (FDTD) simulations suggest the existence of such hot-spots,[17-19, 27] linear light scattering spectra of single nanosponge have so far showed rather broad SP resonances superimposed with some minor spectral modulations.[17, 19, 28] Even for round particles, strong polarization anisotropies were found which are more pronounced in light scattering than in photoluminescence.[19, 28] This has been taken as an indirect signature of light localization. Recent ultrafast photoemission experiments suggest the existence of SP modes with lifetime longer than 20 fs and indicate that Fano-type interferences between localized and delocalized SP modes contribute to the scattering of far-field light.[17] Yet, so far, all these far-field experiments, even if performed on a single porous particle, average over an ensemble of modes, and studies of individual localized SP modes (hot-spots) are lacking.

Here we use scattering type scanning near-field optical microscopy (s-SNOM) to directly probe SP modes on the surface of an individual nanosponge. The hot-spot modes are found to localize on a 10 nm scale and display linewidths of 20 nm or even less. The measurements predict Purcell factors of single hot-spot modes up to $10^6$. Near-field intensity fluctuations with strongly non-Gaussian statistics demonstrate strong SP localization. Our results make nano-sponges an intriguing example of a quasi-3D disordered medium that concentrates far-field light into a few hot-spots with properties that can be controlled by tailoring the nanosponge geometry.

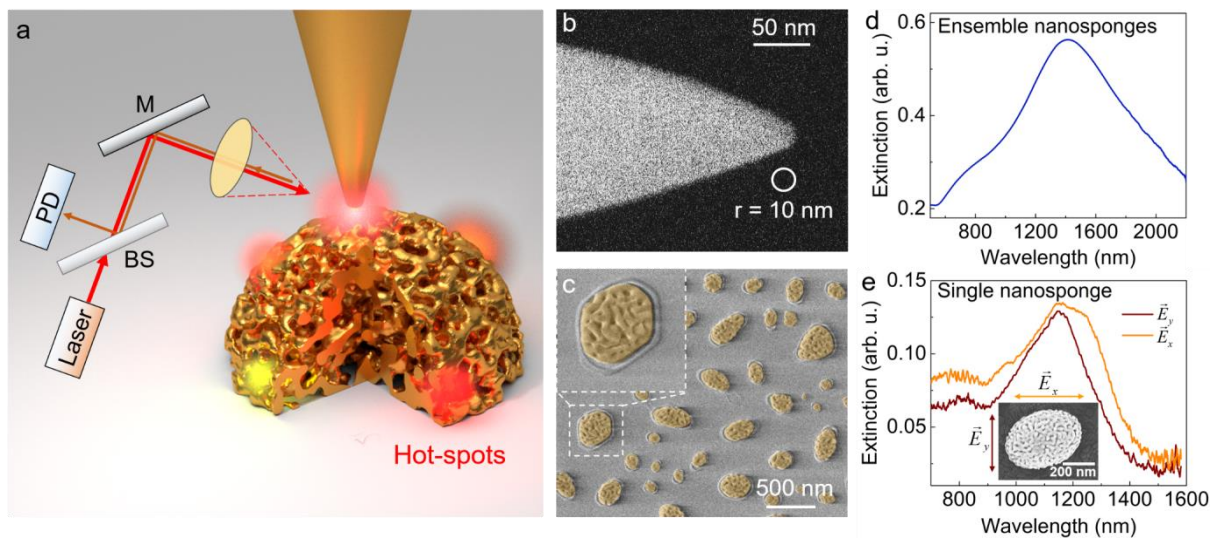

**Figure 1**. (a) Schematic of the scattering-type near-field spectroscopy experiment probing surface plasmon (SP) localization in individual gold nanosponges. Multiple coherent scattering of SP within the disordered, percolated sponge structure leads to the formation of localized modes ("hot-spots"). (b) SEM image of a gold taper with a radius of 10 nm used as near-field probe. (c) False-colour SEM image of gold nanosponges deposited on a glass substrate. (d) Far-field extinction spectrum of a large ensemble of nanosponges shown in (c). (e) Polarization-dependent far-field extinction spectra of an individual nanosponge, shown in the SEM image in the inset. The direction of the polarization relative to particle axis is indicated in the figure. Note the different wavelength scales in (d) and (e).

In this work, we investigate nanoporous gold particles, nanosponges, fabricated by dewetting and dealloying of a gold-silver bilayer.[13-15] The nanosponges are deposited on a glass substrate and have a half-spherical or semi-ellipsoidal shape. An SEM image in Fig. 1c displays an imhomogeneous distribution of particles with different shape and size (150-400 nm diameter). All particles are perforated by a randomly distributed nanopores with a typical diameter of 10-20 nm. A cross-sectional analysis proves that these nanopores are percolating not just the surface but the entire interior of the particle, forming a continuous quasi-3D ligament network.[13, 17, 19] Figure 1d shows the extinction spectrum for a large ensemble of nanosponges from the sample in Fig. 1c. The spectrum shows a broad and weakly structured resonance covering the visible to near infrared region from 600 to 2200 nm. This broad resonance certainly reflects the optical properties of an inhomogeneously broadened ensemble of particles with different geometries. When recording the extinction spectrum of a single nanosponge (Fig. 1e), the spectrum narrows considerably. Even at the single-particle level, however, the spectra show quite broad resonances, covering several hundreds of nm, which are superimposed with several weak sidepeak modulations. The resonance wavelength and modulation amplitude of those sidepeaks vary substantially, even for particles with similar size and porosity, reflecting the unique randomly disordered structure of each particle.[17, 19] In comparison to particles of similar shape but without pores, the spectra display a red shift of the resonance,[13, 15] more pronounced sidepeak modulations and a strong polarization anisotropy (Fig. 1e), observed even for almost spherical particles.[19, 28] The sidepeak modulation has been a taken as a signature of plasmon localization and hot-spot formation, but the linewidth of those sidepeaks is much broader than what is expected from the long hot-spot lifetimes (> 20 fs) deduced from recent time-resolved photoemission experiments.[17] This might already indicate that even at the single particle level, the sidepeak modulations reflect an average over the light scattering contributions from several localized modes.

To demonstrate the existence of localized plasmonic modes in the nanosponges and to analyze the optical properties of single hot-spots in real space, we used s-SNOM to record spatially-resolved scattering spectra of individual nanosponges. In contrast to electron-beam based spectroscopies such as electron energy loss spectroscopy (EELS)[29, 30] or cathodoluminescence (CL)[31], s-SNOM allows us to resonantly excite plasmonic modes with high spatial (~ 10 nm) and spectral (< 10 meV) resolution. A schematic of our s-SNOM experiment is sketched in Fig. 1a. Electrochemically etched, single crystalline gold tips[32] with a radius of curvature of $r \approx 10$ nm are used as scattering probes (Fig. 1b). The distance between tip and sample, $z = z_0 + A\sin(\omega t)$, is controlled by tapping mode atomic force microscopy (AFM), with a tapping frequency $\omega = 2\pi f$ ($f \approx 25$ kHz) and amplitude $A = 5$ nm. During the measurements, the average tip-sample distance $z_0$ is set to 5 nm. By scanning the sample, we simultaneously obtain the topography and the optical scattering signal of the nanosponges. The sample is excited from side by a tunable, continuous-wave Ti:sapphire laser through a 20× (NA = 0.35) objective with a focus diameter of about 1.5 µm. The back-scattered light is collected by the same objective and detected with an avalanche photodiode (APD). When the tip approaches the sample, the optical near-field of the externally illuminated tip is coupled to the optical modes of the sample.[33-37] This tip-sample interaction enhances the local light scattering and results in the emission of a near-field $E_{NF}$ signal from the apex region when the tip is close to the surface. In addition, there is also a spatially much less localized background field, $E_B$, due to the back-reflection of the incident laser from the tip-sample region. In general, several reflection pathways may contribute to this background (see Fig. S1, Supporting Information). In the case of the strongly scattering gold taper, $E_B$ is largely dominated by strong reflection of the focused laser off the taper shaft.[38] We take this reflected field $E_R$ as a reference for the weak near-field scattering $E_{NF}$. The intensity of all interfering scattering contributions is measured with the APD and demodulated at the *n*-th harmonic of the tapping frequency *f* with a lock-in amplifier.[33, 35] In our experiment we simultaneously record harmonic orders $n = 1-4$. For sufficiently high harmonic order, the scattering signal at the sample position $\vec{r}$ and average tip-sample distance $z_0$ is $I^{(n)}(\vec{r}, z_0) = \text{Re}(E_R E_{NF}^{(n)*})$, i.e., it is dominated by the interference between the constant reference field $E_R$ and the local *n*-th harmonic near-field $E_{NF}^{(n)}(\vec{r}, z_0) \propto E_{NF,0}(\vec{r}) \cdot c^{(n)} \cdot e^{-\frac{z_0}{L}}$. Here $E_{NF,0}(\vec{r})$ is the near-field at $z = 0$ and $c^{(n)}$ the Fourier coefficient of the near-field signal at the demodulation order $n$.[33, 35] We assume a near-field contribution $E_{NF}(\vec{r}, z) = E_{NF,0}(\vec{r}) \cdot e^{-\frac{z}{L}}$, which is characterized by an exponential decay of length $L$. This near-field decays much more quickly with tip-sample distance than all background fields. Therefore, the contributions of the background to high order harmonics ($n = 3, 4$) can be safely neglected (see Fig. S1, Supporting Information), ensuring the probing of the local near-field only.

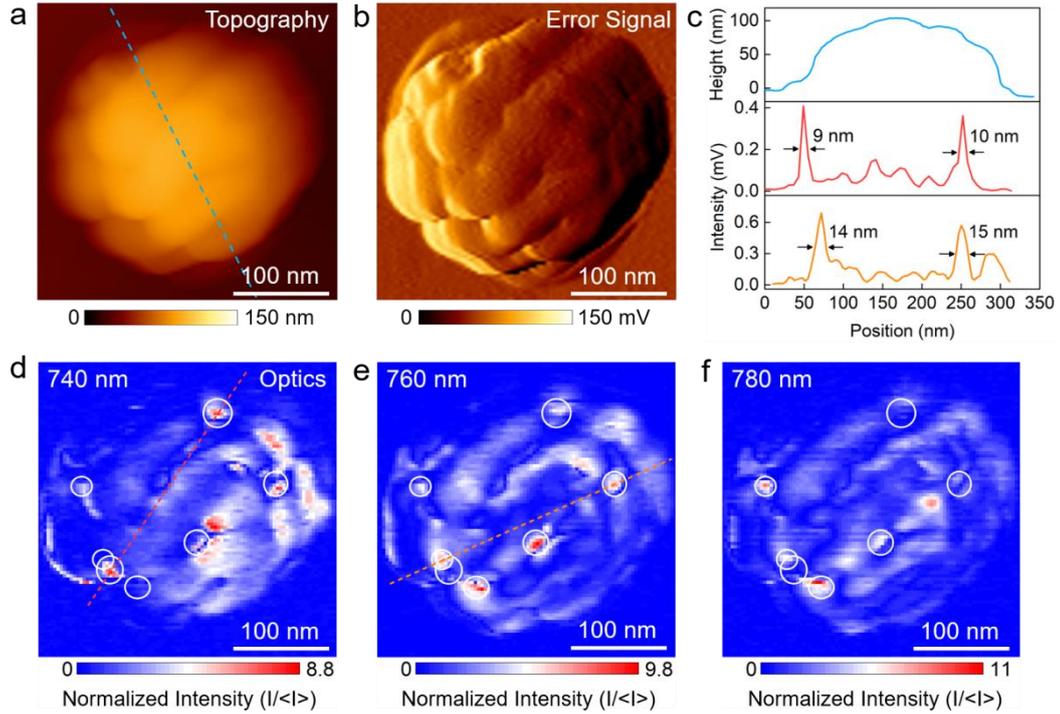

**Figure 2.** (a) AFM topographic image of a single nanosponge. (b) Simultaneously acquired AFM error signal highlighting the pore structure of the sponge. (c) Top: Cross section of the sponge topography along the dashed line in panel (a). Bottom: Cross sections of the scattering signal along the dashed lines in panels (d) and (e), respectively. The 3*f*-signal, recorded at the 3$^{rd}$ harmonic of the tip modulation frequency *f*, reveals SP localization in individual hot-spots with diameters of about 10 nm. (d-f) Optical near-field scattering image (3*f*-signal) of the same nanosponge for laser excitation at 740, 760, and 780 nm, respectively. Several randomly distributed hot-spots are marked with white circles in (d-f).

Figure 2a and 2b show representative AFM topography and error signal images from a single nanosponge, respectively. This nanosponge has a semi-spherical shape with a diameter of about 250 nm and a maximum height of 100 nm (Fig. 2c). The pores are faintly seen in the topography and are much more obvious in the error signal. Figure 2d-f show the optical images of 3*f*-signal of the nanosponge excited at laser wavelengths of 740 nm (d), 760 nm (e), and 780 nm (f), respectively. In all images, we can readily observe localized regions of high near-field amplitude. These "hot-spots" have typical diameters of 15 nm or less and are apparently randomly distributed across the surface. Cross sections of the scattering signal along the dashed lines in Fig. 2d and 2e reveal a spatial localization down to 10 nm (Fig. 2c). Generally, we find localization lengths that are smaller than 15 nm (full width at half maximum, FWHM). These hot-spots appear only at certain resonance wavelengths, as can be seen for selected spots marked by white circles in Fig. 2d-f. When excited at different laser wavelengths, the amplitude of each hot-spot varies significantly while the spatial position remains the same. From these images we see that even a small change of wavelength of 20 nm leads to dramatic changes of the field distribution. This suggests that the resonances of these hot-spots are quite narrow, as will be shown later in detail. In addition to this hot-spot emission, we also observe more delocalized regions with sizeable near-field amplitude. For these delocalized modes, the scattering signal is considerably smaller than that of the hot-spots and their spatial distribution appears to correlate with the nanopore structures seen in the error signal in Fig. 2b. A series of near-field images of this particle recorded in a broader wavelength range from 720 nm to 840 nm (Fig. S2) is compared to representative near-field optical images of several other particles (Fig. S4-5 in the Supporting Information). For all particles that we have studied, we find that the near-field scattering signal is dominated by several spatially and spectrally highly localized hot-spots that are randomly distributed across the nanosponge surface.

The results in Fig. 2 nicely show the very high spatial localization of the hot-spot modes at the surface of our nanosponges. Our s-SNOM demodulation technique also provides information about field localization in surface normal direction (Fig. 3a).[39] For such highly localized modes, the characteristic in-plane wave vector component is much larger than the free-space wave vector and thus the decay length of the optical near-field along the surface normal is on the order of the tapping amplitude (~ 10 nm) or even shorter. Hence, during its periodic oscillation, the tip probes a rapidly decaying local optical near-field (Fig. 3b). The nonlinear tip-sample distance dependence of the near-field amplitude results in an anharmonic periodic oscillation of the scattering signal, $I = \mathrm{Re}(E_R E_{NF}^*)$ (Fig.3b, top inset), giving rise to high-order harmonics in the lock-in output (Fig. 3b, bottom inset). For a typical exponential decay length $L$ of the electric near-field, the demodulated signal at the $n$-th harmonic order is proportional to the Fourier coefficients $c^{(n)} = \frac{1}{T}\int_0^T e^{-\frac{A}{L}\sin(\omega t)} \cdot e^{-in\omega t} dt = (-1)^n \cdot I_n\left(\frac{A}{L}\right)$. Here, the oscillation period $T = 1/f$, and $I_n$ is a modified Bessel functions of the first kind and of order $n$.[35] Via the latter, the Fourier coefficients depend on the near-field decay length $L$.

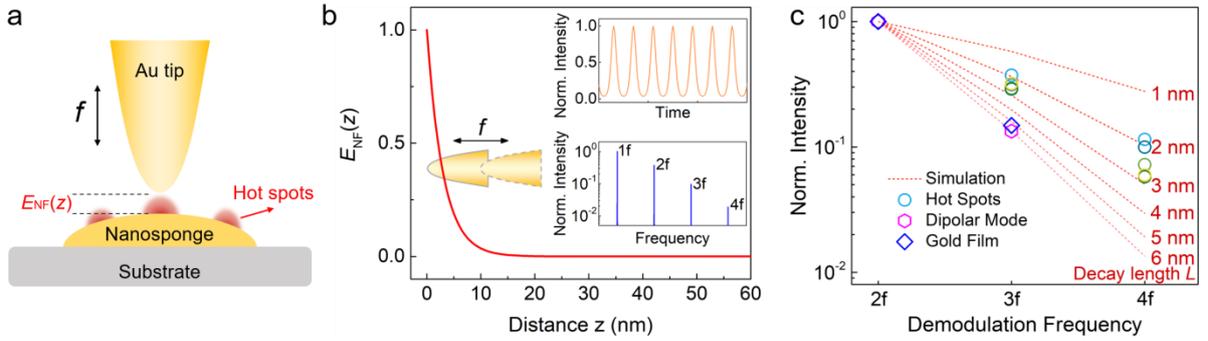

**Figure 3**. (a) Schematic of the experiment probing the decay length of localized hot-spot modes along the surface normal. The tip is modulated at frequency $f$ and scattering signals are simultaneously recorded at the first four harmonics. (b) Simulation of the electric near-field decay, $E_{NF}(z)$, along the surface normal ($L$= 3 nm), probed by a tapping tip. Inset: (Top) Time variation of the near-field intensity during tip modulation for an average tip sample-distance $z_0$ = 5 nm and a tapping amplitude $A$ = 5 nm. (Bottom) Fourier-transformed scattering intensity at harmonic order $n = 1 - 4$. (c) Scattered near-field intensity at second to fourth harmonics from several hot-spots (open circles), from a nanosponge region outside the hot-spots (open hexagons), and from a planer gold surface (open rectangles). Simulated signals for near-field modes with different decay lengths are plotted as dashed lines. All data are normalized to the respective 2*f*-intensities. An average decay length of the hot-spot modes of 2-3 nm is deduced.

We plot the 2*f*-4*f* intensities measured for several hot-spots as open circles in Fig. 3c. For a better comparison, the data are normalized to the respective 2*f*-intensities. We compare these results to related data recorded in a nanosponge region outside the hot-spots (open hexagons) and from a planar gold film surface (open rectangles). Note that for the latter two data sets, the $4f$-signal was too weak to be resolved. To deduce the near-field decay length, we calculated the scattered intensity $I = \text{Re}(E_R E_{NF}^*)$ by assuming a purely exponential near-field decay (Fig. 3b), and extracted the $nf$-intensities from Fourier-transformed spectra (Fig. 3b, bottom inset). The simulated results with decay length of $L = 1$-$6$ nm are plotted in Fig. 3c as dashed lines.

By comparing the experimental and simulated data, we deduce $L \sim 5$ nm for the planar gold film. This decay length matches to that obtained from an optical approach curve (Fig. S8). For such a planar film, tip-sample coupling induces an image dipole in the film and enhances the effective polarizability of the near-field taper.[33, 34] Hence, the measured decay mainly probes the decay of the localized optical near-field at the taper apex. The short decay length seen in both Fig. 3c and in the approach curve (Fig. S8) reflects the small apex diameter of our tips. A similar decay length is seen outside the hot-spots of the nanosponge. This suggests that also here the off-resonant coupling between tip and sample dominates over the resonant excitation of localized hot-spots. In contrast, in the hot-spot regions, we see a clearly reduced decay length of only $L \sim 2$-$3$ nm. Here, the scattering signal mostly arises from the scattering of the localized hot-spot fields by the taper apex. Hence, the shorter decay length measures the decay of the hot-spot modes rather than merely that of the tip's near-field. We thus take the deduced decay length of $L \sim 2$-$3$ nm as an estimate of the actual decay length of the hot-spot near-fields. A quantitative analysis[39] may be rather involved since in our simplified model we assume a purely exponential near-field decay and neglect the finite extent of the taper field as well as possible coupling-induced modifications of the hot-spot spectra. Nevertheless, we note that we observed strong signal intensity at the 4$^{th}$ harmonic even for small tapping amplitudes of only 5 nm (Fig. S3 and S6). This can be only understood if the hot-spot decay length is shorter than 5 nm.

The SNOM measurements in Fig. 2 suggest very efficient coupling of tip dipole to localized hot-spot modes. In these experiments, the incident light is *p*-polarized, i.e., polarized mainly along the taper axis. This induces a strong, localized near-field at the apex with a dimension and polarization that is well matched to that of the localized modes. This mode matching makes coupling very efficient. Recent photoemission experiments suggest that also the excitation of the collective dipole mode of the particle contributes to laser-triggered photoemission.[17] The SNOM measurements at high harmonic order (Fig. 2d-f) provide little evidence on this collective dipolar mode. In data recorded at lower harmonics (Fig. S3), however, their excitation is seen more clearly. Since this mode is more delocalized, the decay along the normal direction is weaker than that of the hot-spot modes. Hence the contribution of the collective modes to the higher order harmonic SNOM images is reduced. So far, it is difficult to quantify the coupling between collective and hot-spot modes in our nanosponges. For this, more advanced nonlinear optical experiments seem necessary.

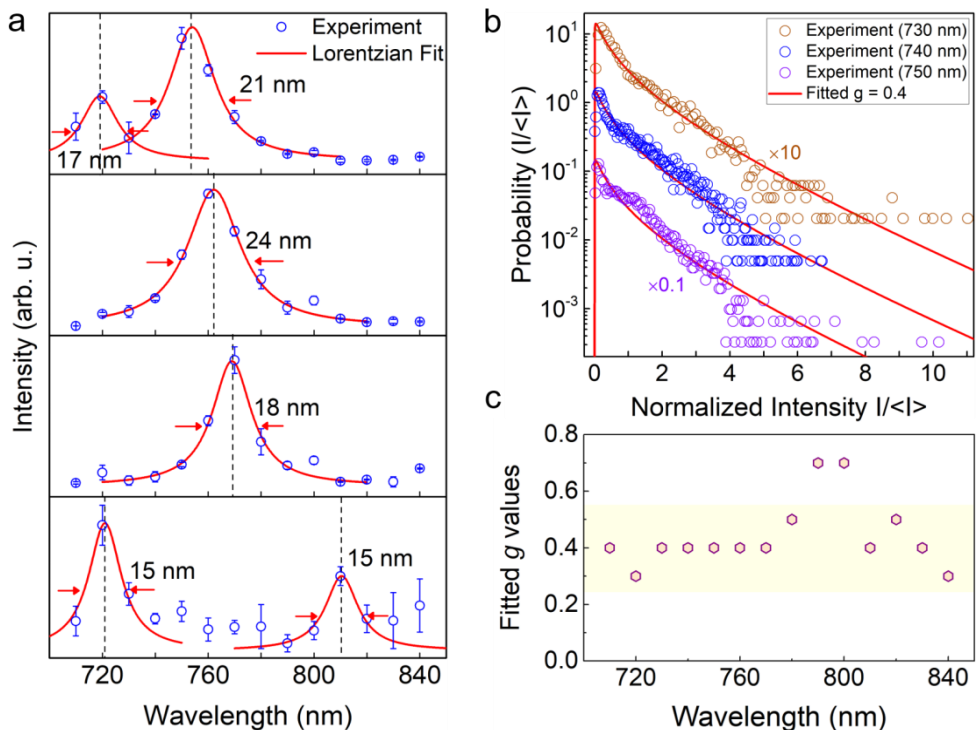

**Figure 4**. (a) Near-field scattering spectra of individual hot-spots, recorded at the 3$^{rd}$ harmonic of the tip modulation (open circles). The red lines show resonances with Lorentzian line shape,

with linewidths of only 15 - 25 nm. The quality factor of those hot-spots can exceed 40. (b) Histograms of the 3*f*-scattering intensity from a single nanosponge, recorded for three excitation wavelengths. The data are normalized to the average scattering intensity and are vertically shifted for clarity. Strongly non-Gaussian statistics reveal pronounced fluctuations of the local near-field intensity. The red lines show fits based on a single parameter scaling model.[40] (c) Wavelength dependence of the scaling parameter *g* characterizing the near-field fluctuations. The values are deduced from fits to histograms as shown in (b). At all wavelengths, the SP modes of the nanosponge are in the strong localization (*g*<*1*) regime.

We now discuss local near-field spectra obtained from a series of wavelength-dependent scans such as those shown in Fig. 2d-f. The simultaneously recorded error signals (Fig. 2b) allow us to track the evolution of individual hot-spots with high precision. Representative hot-spot spectra are shown in Fig. 4a. Two remarkable features can be observed from these spectra. First, we see sharp resonances with narrow linewidth down to $\delta\omega \sim 20$ nm (FWHM). Similarly sharp resonances are also seen in the scattering spectra of hot-spots from another sponge, shown in Fig. S7. All peaks can reasonably be well fitted by a Lorentzian line shape (red lines) confirming the narrow linewidths and indicating high quality factors, *Q*, exceeding 40. The *Q* factors of these localized hot-spot modes are thus significantly higher than those of previously studied gap plasmon cavities (~10-15).[21] Second, all hot-spot resonances are spatially (Fig. 2c) and spectrally (Fig. 3a) well separated and we did not observe local spectra with overlapping resonances. This level repulsion is a signature of random mode localization.[41, 42] In most spectra, we observe only one eigenmode in the detection range. For some hot-spots, however, two eigenmodes are observed and the modes are spectrally separated by a spacing $\Delta\omega$ of at least 35 nm (Fig. 4a, top spectrum). This mode spacing is larger than the average linewidth $\delta\omega$. In the theory of random mode localization, the ratio between linewidth and mode spacing, $\delta = \delta\omega/\Delta\omega$, the Thouless number, is a fundamental localization parameter.[42] A $\delta < 1$ means that the plasmonic modes are strongly localized and have weakly overlapping mode profiles.[42]

We find $\delta < 0.57$ in all our spectra, suggesting that the plasmonic modes in nanosponges is indeed in the strong localization regime.

To get a statistically significant measure for the mode localization, we now analyse the spatial intensity fluctuations that are seen in the near-field optical images in Fig. 2. For this, we normalize the local scattering intensity, $I$, in each image to the spatially averaged intensity $<I>$. Representative histograms recorded from a single nanosponge (Fig. 2) at different excitation wavelengths are shown in Fig. 4b. In all histograms, we find marked deviations from Gaussian statistics with a long, slowly decaying tail at large $I/<I>$. Such non-Gaussian statistics have been taken as a signature of strong, Anderson-type localization of optical modes.[43-46] This long tail is dominated by localized modes with strong field amplitudes. To analyse our histograms, we compare them to a single scaling parameter model for light transport through multiply scattering media introduced by Nieuwenhuizen *et al.* in Ref. 40. This model has been used successfully, e.g., to characterize mode localization in randomly disordered zinc oxide nanoneedles[47] and quasi-two-dimensional waveguides.[45, 46] We find that the data in Fig. 4b are reasonably well described by the Nieuwenhuizen model with a single scaling parameter, $g = 0.4$.[40] For microwave transport through random media it has been shown that this dimensionless conductance ($g$) is equal to the Thouless number.[42] Similarly small $g$ values are found for all excitation colours that we tested (Fig. 4c). Hence, this statistical analysis firmly supports our conclusion that the plasmonic excitations of our nanosponge samples are in the strong localization regime ($g < 1$). Such a strong localization results from the multiple coherent scattering of plasmons within the randomly disordered nanoporous network.

Evidently, this scattering results in the formation of spatially highly-localized hot-spot modes with high $Q$ values, suggesting long lifetimes and low damping rates of those modes. For a homogeneously broadened Lorentzian resonance, describing well our measured hot-spot spectra, the FWHM linewidth $\Gamma = 2\hbar/T_2$ is directly proportional to the total dephasing time $T_2$ of the transition.[16, 48] From the experimental spectra of several localized modes (Fig. 4a), the linewidth $\Gamma$ = 30~56 meV corresponds to a dephasing time of the localized modes of $T_2$ = 24~44 fs. Since for plasmonic systems pure dephasing process are negligible,[16, 48] the lifetime of the plasmonic mode, $T_1 = T_2/2$, is ranging from 12 to 22 fs. These values agree with mode lifetimes of ~ 20 fs estimated from recent time-resolved photoemission measurements probing inhomogeneously broadened ensembles of modes in a single nanosponge.[17] The lifetime of localized hot-spot modes in nanosponges is therefore much longer than common lifetimes of plasmonic nanoparticles of less than 5 fs.[48, 49]

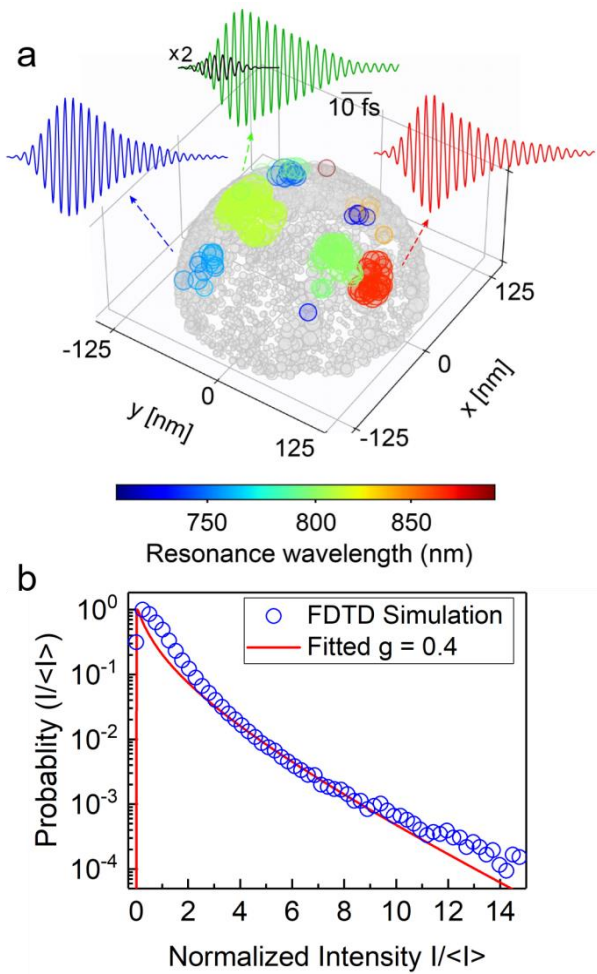
15

**Figure 5.** Calculated amplitude and resonance wavelength of the localized modes (colour circles) at the surface of a nanosponge. The diameter and colour of the circles represent the amplitude and wavelength of the dominating modes, respectively. Hot-spots with different resonance wavelengths are randomly distributed across the surface. Insets: simulated electric field dynamics of three representative hot-spot modes together with the 6-fs excitation pulse (black line). A 6-8 times field enhancement in the hot-spots is observed. (b) Histograms of the calculated near-field intensity distribution (open circles) in (a) and a fit to single parameter scaling model with a dimensionless conductance of $g = 0.4$ (red line).

To gain more insights into the plasmon excitation and field dynamics, we simulated the time-dependent local electromagnetic fields at the surface of a nanosponge.[17] A nanosponge model is created by using a golden half-sphere of 250 nm diameter, percolated with 20 nm nanopores (Fig. 1a, see details in Methods). The sponge is optically excited from the side by a pulsed, p-polarized plane wave source, incident at 20° with respect to the substrate surface, consistent with the experimental conditions. The source has a Gaussian time profile (Fig. 5a, inset: black line) with a pulse duration of 6 fs (FWHM of the intensity profile $I(t) = |E(t)|^2$). Figure 5a shows the frequency-resolved spatial distribution of the amplitudes of the local modes at the surface of the nanosponge, at a time delay of 40 fs after the arrival of the excitation pulse. The field amplitude is obtained from a spectral analysis by harmonic inversion (see Supporting Information). For such comparatively long time delays, only long-lived modes can be observed in Fig. 5a. Several localized modes (colour circles) with different resonance wavelengths are found to be randomly distributed across the surface of nanosponge. In general, these findings reproduce well the experimental results presented in Fig. 2. An analysis of the electric field dynamics of the hot-spot modes shows a coherent excitation persisting well over 40 fs (Fig. 5a, insets), much longer than the duration of the excitation pulse (black line). The simulated field dynamics is in good agreement with the dephasing times of 24~44 fs derived from experiment. Importantly, we observe a 6-8 times field enhancement of hot-spot modes compared to the incident field. The enhancement is deduced by comparing the maximum field amplitude of the local near-field to that of the excitation pulse. The simulations clearly illustrate the pronounced field localization and enhancement of the hot-spot modes resulting from the multiple coherent scattering of SP within the randomly disordered structure. We further analysed the intensity fluctuations of the local near-fields that are seen in the simulations shown in Fig. 5a. The resulting histogram (Fig. 5b) displays non-Gaussian statistics similar to those seen in the experimental histograms. The simulated data are reasonably well reproduced by the single parameter scaling when taking a dimensionless conductance of $g = 0.4$.[40] This is an additional convincing support for the strong localization of plasmonic excitations in our nanosponges. Hence, the simulation results fully corroborate our experimental observation of highly spatially localized, long-lived hot-spot modes at the surface of percolated nanosponges.

The long lifetimes of the localized modes mean that both radiative and non-radiative losses of the hot-spot modes are small. The weak radiative damping evidently reflects the small mode volumes of hot-spots. Nevertheless, in contrast to a metal particle of the same size, e.g., a 10-nm diameter gold sphere with 2.5 fs lifetime,[48] the lifetime of hot-spot modes is enhanced by almost an order of magnitude. Hence, the non-radiative damping due to the excitation of electron-hole pairs is also much reduced. This likely means that most of the electromagnetic energy is confined outside the metal in our large surface-to-volume ratio nanosponges.

The narrow linewidth of the localized plasmon modes and their strong spatial confinement make them excellent candidates for exploring the coupling of quantum emitter (QE) to these plasmon nanocavities.[50] This coupling can substantially alter the optical properties of the QE. In the weak coupling regime, the radiative damping of the QE is enhanced by a Purcell factor $F_P = \frac{3}{4\pi^2}\left(\frac{\lambda}{n_0}\right)^3 \frac{Q}{V_M}$ since the presence of the nanocavity locally increases the optical density of states.[51, 52] In the elusive strong-coupling regime, a periodic transfer of energy between QE and nanocavity is expected.[25] For a single molecular QE coupled to a plasmonic nanocavity this strong-coupling regime has recently been reached in Ref. 21. It requires a coupling strength of the transition dipole moment of the QE to the vacuum field of the cavity that exceeds the damping rate of both the plasmonic nanocavity ($\gamma_{pl}$) and the QE ($\gamma_e$).[20-25, 50] For a number of $N$ QE with transition dipole moment $\mu_e$, the coupling strength is $g = \sqrt{N}\mu_e |E_{vac}|$. Here, the local vacuum field amplitude $E_{vac} = \sqrt{\hbar\omega/(2\varepsilon\varepsilon_0 V_M)}$ scales inversely with the mode volume $V_M$ of the plasmonic nanocavity. For our nanosponges, we estimate a mode volume of (10 × 10 × 6) nm³ = 600 nm³. As the lateral extent of the mode, we take the measured hot-spot diameter of ~10 nm (Fig. 2c). In the normal direction we estimate an extension of 6 nm, given by the sum of decay length along the surface normal (~3 nm) and the field penetration into the sponge of 3 nm. This small mode volume implies a giant local vacuum field amplitude of ~0.15 V/nm at a resonance wavelength of $\lambda$ = 800 nm ($\omega = 2\pi c/\lambda$). These numbers imply that a single, correctly oriented QE with a dipole moment exceeding ~20 D and a linewidth of less than 30 meV would be strongly coupled to the randomly localized plasmonic cavity mode. For emitters with comparatively large dipole moments (e.g., J-aggregated dye molecules, $\mu_e$ ~100 D[53, 54]) that are placed inside these fields, this would results in dipolar coupling constants as large as 310 meV, exceeding by far the linewidth of the plasmonic mode ($\Gamma$ ~40 meV). This unique combination of high $Q$ and low $V_M$ makes the randomly disordered nanocavities in nanosponges extremely attractive for various types of strong-coupling studies. Even in the weak coupling regime, the presence of the cavity would have a profound impact on the dynamics of the QE. Taking an effective refractive index $n_0$ = 1, the measured values of the quality factor $Q$ = 40 and the mode volume $V_M$ = 600 nm³, we predict a Purcell factor as large as $F_P$ = 2.5 × $10^6$, matching record values estimated for the best gap plasmon resonators that have been designed so far.[21] This would imply, for instance, that the radiative decay time of a QE with a spontaneous emission time of 10 ns is reduced by a factor of $10^6$ to a mere 10 fs. To our knowledge such huge Purcell effects have not yet been demonstrated experimentally in a solid state system.

To summarize, we directly verified the existence of highly spatially and spectrally localized plasmonic modes at the surface of individual gold nanosponges, percolated with a three-dimensional ligament network, by near-field spectroscopy. These modes localize on a 10 nm length scale and have ~20 nm linewidth, resulting in exceptionally high Purcell factors on the order of $10^6$. Our findings demonstrate nanosponges as an intriguing alternative to ordered plasmonic nanostructures, supporting strong plasmon localization in a broad spectral range that can be tailored by varying the geometry, pore size, filling factor, and/or composition of the particle. We find that nanosponges not only support a high density of localized hot-spot modes but also a collective plasmon excitation with large extinction cross section which couples strongly to far-field light. This turns nanosponge into an efficient nanoantenna which is coupled to a series of randomly localized hot-spots with highly confined and enhanced electric fields. This may result in ultra-efficient nonlinear plasmonic hybrids by infiltration with a variety of nonlinear materials and creates a new platform for exploring strong coupling to quantum emitters in a wide energy range. This certainly warrants more efforts towards the tailored design of these particles. Studies of the coupling of these localized hot-spots to various types of active quantum emitters, using advanced spectroscopy such as two-dimensional optical nanoscopy, are currently underway in our laboratories.

ASSOCIATED CONTENT

**Supporting Information**.

The following files are available free of charge.

Experimental section, more SNOM images and hot-spot spectra on other nanosponges (PDF)

AUTHOR INFORMATION

**Corresponding Author**

*christoph.lienau@uni-oldenburg.de.

**Author Contributions**

C.L., E.R., and P.S. initiated the project. C.L. conceived the experiment. D.W. and P.S. designed and fabricated the samples. J.Z., A.C. and A.K. performed the experiments and analyzed and discussed the data together with C.L. F.S. and E.R. performed the FDTD calculations. J.Z. and C.L. prepared the manuscript.

‡These authors contributed equally.

ACKNOWLEDGMENT

Financial support by the Deutsche Forschungsgemeinschaft (SPP1839 "Tailored Disorder", grants LI 580/12, RU 1383/5, SCHA 632/24, SPP1840), the Korea Foundation for International Cooperation of Science and Technology (Global Research Laboratory project, K20815000003) and the German-Israeli Foundation (GIF grant no. 1256) is gratefully acknowledged. We thank Germann Hergert for collecting the single particle extinction spectra. J.Z. acknowledges support from Alexander von Humboldt Foundation. A.C. is supported by German Academic Exchange Service (DAAD) for a Ph.D. scholarship. A.K. acknowledges the financial support from the Graduate Program "Nanoenergy Research" of the State of Lower Saxony.

**Notes**

Any additional relevant notes should be placed here.